\documentclass[12pt]{iopart}
\usepackage{graphicx}
\usepackage{dcolumn}
\usepackage{bm}
\usepackage{iopams}
\usepackage{cite}
\usepackage{color}

\begin{document}
\title{Quasielastic $(p,n)$ reactions described by a microscopic optical model based on the Gogny force}
\author{Juan Lopez Mora\~{n}a$^1$ - Xavier Vi\~{n}as$^{1,2}$$^*$}
\address{$^1$ Departament de F\'isica Qu\`antica i Astrof\'isica and Institut de Ci\`encies del Cosmos (ICCUB),
Facultat de F\'isica, Universitat de Barcelona, Mart\'i i Franqu\`es 1, E-08028 Barcelona, Spain\\
$^2$Institut Menorqu\'i d'Estudis, Camí des Castell 28, 07702 Ma\'o, Spain}
$^*$E-mail: xavier@fqa.ub.edu (corresponding author)\\
\date{\today}
\setlength\mathindent{20pt}
\begin{abstract}

In this work we want to study quasielastic $(p,n)$ exchange reactions using a semi-microscopic optical model derived in a previous work \cite{lopez21}
based on a nuclear matter approach where the real and imaginary parts are given by the first and second order terms, respectively, of the mass
operator obtained by a Brueckner-Hartree-Fock calculation using a G-matrix built up with an effective Gogny interaction. The study of these
quasielastic reactions is performed within a Distorted Wave Born Approximation (DWBA) to evaluate the wave functions in the entrance and exit 
channels, which in turn are used to compute the transition matrix elements. 
This model, which is free of adjustable parameters, provide a reasonable good agreement with the considered experimental data, namely differential
cross sections, analyzing powers and total cross sections, of different reactions spanned along the periodic table at several energies.

\end{abstract}

\section{Introduction}
 Almost sixty years ago the analysis of the neutron spectra in $(p,n)$ exchange reactions revealed that these interactions excited 
quite strongly the isobaric analog state (IAS) of the target ground-state \cite{anderson61,anderson62a,anderson62b,anderson63}. Assuming that the
isospin symmetry is exact, the IAS has the same structure as the ground-state of the target, except that a neutron has been replaced by
a proton. In the isospin representation these two states, namely the ground-state of the target and the IAS, belong to the same isospin
multiplet ${\bf T}$ differing only in the third component. If the target has $N$ neutrons and $Z$ protons, its isospin third component
is $T=(N-Z)/2$ while the IAS with an additional proton and a neutron less has $T-1$. In this scenario the energy of the IAS differs
from the one of the ground-state by the Coulomb energy of the added proton. The large similarity between the initial and final states, 
which differs in the flipped isospin of the incident projectile shows that this type of $(p,n)$ reactions can be regarded as elastic
(more precisely quasielastic) scattering process. As a consequence, it seems plausible to explain the transition between the initial and 
final states in terms of an one-body potential in close analogy with the optical model potential for elastic nucleon-nucleus scattering 
\cite{satchler64}.

As it was pointed out by Lane \cite{lane62}, the nuclear part of this optical potential may be written in a charge independent form,
which allows to describe not only proton and neutron elastic scattering from nuclei but also the $(p,n)$ reaction from the ground-state 
of the target to its corresponding IAS. In the Lane picture the central part of this optical potential reads:
\begin{equation}
U(\vec{R})= U_0(\vec{R})+ 4\frac{\vec{t}\cdot\vec{T}}{A}U_1(\vec{R})\nonumber \\
=U_0(\vec{R})+ 4\frac{t_0T_0}{A}U_1(\vec{R})
+4\frac{t_+ T_- + t_- T_+}{2A}U_1(\vec{R}),
\label{lane}
\end{equation}
where ${\bf t}$ and ${\bf T}$ are the isospins of the incident proton and target nucleus, respectively.
The second term of the first equality in Eq.(\ref{lane}) is the so-called Lane potential and contributes to both $(p,p)$ and $(n,n)$ 
elastic scattering (second term of the second equality in (\ref{lane})) as well as to the charge exchange reaction $(p,n)$ \cite{satchler83},
which completely determines the $\Delta J^{\pi}=0^+$ transition strength of the reaction exciting the IAS of the ground-state of the target.
Eq.(\ref{lane}) points out that there are two possible ways to experimentally extract the isovector contribution to the optical potential.
One possibility is to perform measurements of $(p,p)$ and $(n,n)$ elastic scattering observables from the same target at the same energy
in such a way that the difference provides some information about $U_1$. The second way is to study the isospin flip (charge exchange) between 
ground-state of the target and its IAS.

Several measurements of the $(p,n)$ reaction observables at different energies and from different targets have been performed in the past.
Different phenomenological Lane potentials $U_1$ have been fitted to reproduce these experimental data 
\cite{satchler64,anderson64,hansen64,carlson73,carlson75,jolly73,doering75,patterson75,jon97,jon00} providing angular distributions and reaction 
cross sections. Also in \cite{gosset76} the analyzing power in quasielastic $(p,n)$ reactions on several targets at 22.5 MeV have been analyzed. 
From the theoretical side a semimicroscopic Lane  potential based on the Jeukene-Lejeune-Mahaux (JLM) microscopic model 
\cite{jeukene74,jeukene76,jeukene77a,jeukene77b} modified later to reproduce elastic scattering in finite nuclei \cite{bauge98} has been applied 
for the study of $(p,n)$ quasielastic charge exchange reactions \cite{bauge01}. Later on a consistent analysis of charge exchange reactions in 
$^{48}$Ca, $^{90}$Zr, $^{120}$Sn and $^{208}$Pb targets at 35 and 45 MeV is done using the folding model within the two-channel coupling formalism 
\cite{koha07}. Another theoretical study on quasielastic $(p,n)$ charge exchange reactions at high proton energy between 80 and 800 MeV based on an 
extension of a microscopic full-folding optical model for nucleon-nucleus scattering is carried out in Ref.\cite{arellano07}. In a very recent paper 
\cite{whitehead22} the exchange reactions from the ground-state to the isobaric analog state of $^{14}$C, $^{48}$Ca and $^{90}$Zr at several 
bombarding energies have been analyzed using a microscopic optical potential derived from many-body theory with chiral forces in nuclear matter 
\cite{whitehead21} paying special attention to the uncertainties in the differential cross-sections.  
Based on previous optical model analysis of different experimental data  of $(p,n)$ reactions \cite{carlson73,carlson75,jolly73,jon97,jon00}, a very 
interesting paper connecting differential cross sections measurements with the values of the symmetry energy in nuclear  matter and its density 
content has been reported few years ago by Danielewicz and collaborators \cite{danielewicz17}.
      
Recently we derived a microscopical optical model potential (MOP) for describing nucleon-nucleus elastic scattering \cite{lopez21}. Inspired by 
the JLM model, this MOP was built up within a nuclear matter approach, where the real and imaginary parts of the central potential are given by the 
first and second-order terms, respectively, of the mass operator, which was calculated with the Brueckner-Hartree-Fock method using a $G$ matrix 
computed with the effective Gogny interaction. This relatively simple MOP, which does not contain adjustable parameters, provides a reasonable good 
description of experimental data of differential cross sections, analyzing powers and total (neutrons) and reaction (protons) cross sections
 in a range of energies of the projectile up to about 60 MeV for protons and 40 MeV for neutrons, which cover the energies of the
charge exchange reactions that will be analyzed in this work. In this work we call this theoretical model MOP-G from now on.
%To explore the predictive power of this nucleon-nucleus model, we have studied the elastic scattering of light particles, namely $^2$H, $^3$H,<$^3$He 
%and $^4$He, using our MOP within a generalized Watanabe model that finds a description of of the experimental data reasonable \cite{lopez22}. 
In the present work we want to study if the MOP-G is also able to describe quasielastic $(p,n)$ charge exchange reactions. To this end we start 
from the two-channel coupling formalism (see e.g. \cite{koha07}) but without including self-consistency as in the calculations reported in 
\cite{danielewicz17}. This implies that the wave functions in the entrance and exit channels are evaluated by means of the distorted wave Born 
approximation (DWBA). As it is pointed out in \cite{danielewicz17}, the DWBA can be used when the final state is basically populated by one-step 
process, without intermediate states or multiple steps between the initial and final states. The DWBA is appropriate for describing fast processes 
with low transition probability, as for example peripheral reactions at high energy. The quasielastic $(p,n)$ reactions that we will
study in this work meet, actually, these conditions and therefore DWBA is a suitable method to deal with these processes. The paper is organized 
as follows. In the second section the basic theory is presented. In the third section results concerning differential cross sections, reaction cross
 sections and analyzing powers for quasielastic $(p,n)$ charge exchange reactions are discussed. Finally our conclusions are laid in the last 
section.    
 
 \section{Basic theory of the quasielastic $(p,n)$ charge exchange reactions}

Following the  Appendix of Ref.\cite{lane62}, the scattering of incident protons in $(p,n)$ reactions can be described by means of a Hamiltonian 
containing the kinetic energy and the central optical model potential given by Eq.(\ref{lane}) supplemented by the Coulomb potential $V_c(\vec{R})$. 
The corresponding Schr\"odinger equation reads:
\begin{equation}
\bigg( K +  U_0(\vec{R})+ 4\frac{\vec{t}\cdot\vec{T}}{A}U_1(\vec{R}) + (\frac{1}{2} - t_3)V_c(\vec{R})\bigg)\Psi 
= \big(E - (\frac{1}{2} + t_3)\Delta_c\big)\Psi, 
\label{sch}
\end{equation}
where $\Delta_c$ is the Coulomb displacement energy ($\Delta_c=-Q$ for $(p,n)$ reactions in IAS) .

The total wave function $\Psi$ solution of Eq.(\ref{sch}) may be written as \cite{lane62} 
\begin{equation}
\Psi = \vert pA \rangle \chi_{pA}({\bf R}) + \vert n\tilde{A }\rangle \chi_{n\tilde{A}}({\bf R}),
\label{wavef}
\end{equation}
where the waves $\chi_{pA}({\bf R})$ and $\chi_{n\tilde{A}}({\bf R})$ describe the relative motion of the scattering system.
In this equation $|pA>$ and  $|n\tilde{A}>$ indicate the isospin states formed by adding a proton to the target $A$ and a neutron to the 
IAS $\tilde{A}$. These isospin states can be written as superpositions of states of well defined total isospin ${\bf T \otimes t}$ \cite{satchler83}
 as follows:
\begin{eqnarray}
\vert pA \rangle = \frac{1}{\sqrt{2T+1}}\vert T + \frac{1}{2}, T_3 - \frac{1}{2} \rangle
+ \sqrt{\frac{2T}{2T+1}} \vert T - \frac{1}{2}, T_3 - \frac{1}{2} \rangle  \\
\vert n\tilde{A }\rangle = \sqrt{\frac{2T}{2T+1}} \vert T + \frac{1}{2}, T_3 - \frac{1}{2} \rangle
-  \frac{1}{\sqrt{2T+1}}\vert T - \frac{1}{2}, T_3 - \frac{1}{2} \rangle, 
\label{isospinstates}
\end{eqnarray}
and therefore
%\begin{eqnarray}
%&&<pA \vert{\bf t}\cdot{\bf T}\vert pA> = 
%\langle \frac{1}{2}, -\frac{1}{2}\vert \langle T,T \vert{\bf t}\cdot{\bf T}\vert \frac{1}{2}, -\frac{1}{2} \rangle \vert T,T \rangle 
%= - \frac{T}{2} \nonumber \\
%&&<n\tilde{A} \vert{\bf t}\cdot{\bf T}\vert n\tilde{A}> =
%\langle \frac{1}{2}, +\frac{1}{2}; T,T-1 \vert{\bf t}\cdot{\bf T}\vert \frac{1}{2}, +\frac{1}{2}; T,T-1 \rangle = \frac{T-1}{2} \nonumber \\
%&&<n\tilde{A} \vert{\bf t}\cdot{\bf T}\vert pA> =
%\langle \frac{1}{2}, +\frac{1}{2}, T,T-1 \vert{\bf t}\cdot{\bf T}\vert \frac{1}{2}, -\frac{1}{2}; T,T \rangle = \sqrt{\frac{T}{2}}, 
%\label{matrixel}
%\end{eqnarray}
\begin{equation}
<pA \vert{\bf t}\cdot{\bf T}\vert pA> =  - \frac{T}{2}, \quad <n\tilde{A} \vert{\bf t}\cdot{\bf T}\vert n\tilde{A}> = \frac{T-1}{2},
\quad <n\tilde{A} \vert{\bf t}\cdot{\bf T}\vert pA> =  \sqrt{\frac{T}{2}}.
\label{matrixel}
\end{equation}
From these Eqs.(\ref{matrixel})
we can see that the diagonal elements of ${\bf t}\cdot{\bf T}$ shift the proton and neutron optical potentials down and up by a quantity that depend 
on the asymmetry of the target $N-Z$ while the off-diagonal part of ${\bf t}\cdot{\bf T}$ induces transitions from the target ground-state 
$\vert pA >$ to the IAS state $\vert n\tilde{A} >$ in the residual nucleus. This behaviour splits the Schr\"odinger equation (\ref{sch}) acting on 
the wave function  (\ref{wavef}) in two coupled equations:
\begin{equation}
\bigg(K + U_0 - \frac{2T}{A}U_1 + V_c - E_{pA}\bigg)\chi_{pA} + \frac{2\sqrt{2T}}{A}U_1\chi_{n\tilde{A}} = 0 \\
\label{coupledeqp}
\end{equation}
\begin{equation}
\bigg(K + U_0 + \frac{2(T-1)}{A}U_1 - E_{pA} + \Delta_c\bigg)\chi_{n\tilde{A}} + \frac{2\sqrt{2T}}{A}U_1\chi_{pA} = 0,
\label{coupledeqn}
\end{equation} 
where the first terms of these equations (\ref{coupledeqp}) and (\ref{coupledeqn}) represents the $(p,p)$ and $(n,n)$ scattering and the second  
terms $(p,n)$ corresponding to the exchange reactions. Due to the fact that the coupling term is small compared with the elastic scattering 
contributions (the reaction cross sections in the elastic entrance $(p,p)$ and exit channels $(n,n)$ are, at least, two orders of
magnitude larger than in the charge exchange channel $(p,n)$), it seems reasonable to use the DWBA to obtain the $(p,n)$ transition 
amplitude as we mentioned before. To this end, we neglect the second terms in Eqs.~(\ref{coupledeqp}) and (\ref{coupledeqn}) and solve the 
corresponding homogeneous equations, which describe the proton-target and neutron-analog elastic scattering  at energies $E_{pA}$ and 
$E_{n\tilde{A}}=E_{pA} -\Delta_c$, respectively. Once the distorted waves $\chi_{pA}$ and $\chi_{n\tilde{A}}$ are determined (see below), they are 
used to obtain the $(p,n)$ transition amplitude as
\begin{equation}
T_{pn} = - \langle \chi_{n\tilde{A}} \vert \frac{2\sqrt{2T}}{A}U_1 \vert \chi_{pA} \rangle.
\label{transition}
\end{equation}

To study the ability of our microscopic optical potential based on the Gogny force, which is explained in detail in Ref.\cite{lopez21}, we build up 
the isoscalar and isovector parts of the Lane potential starting from the entrance proton and exit neutron optical potentials, which allows to 
write \cite{danielewicz17}
\begin{eqnarray}
U_0({\bf R}) = U_p({\bf R}) + \frac{N-Z}{A} U_1({\bf R}) 
\label{ompot0}\\
U_1({\bf R}) = \frac{A}{2(N-Z-1)}\big[ U_n({\bf R}) - U_p({\bf R})\big] 
\label{ompot1}
\end{eqnarray}    
where $U_p({\bf R})$ and $U_n({\bf R})$ are the proton-nucleus and neutron-nucleus optical potentials in the $A(N,Z)$ and $A(N-1,Z+1)$ targets at 
energies $E$ and $E-\Delta_c$, respectively \cite{danielewicz17}.  

The unpolarized $(p,n)$ differential cross sections in the DWBA are obtained from the transition amplitude as \cite{satchler64} 
\begin{equation}
\frac{d \sigma_(p,n)}{d \Omega} = \bigg(\frac{\mu}{2 \pi \hbar^2}\bigg)^2 \frac{k_n}{k_p} \vert T_{pn}\vert^2.
\label{cs}  
\end{equation}

The distorted waves $\chi_{pA}$ and $\chi_{n\tilde{A}}$, which determine the transition amplitude, are solved as follows. In the optical model 
approach these wave functions are solution of the Schr\"odinger equation:
\begin{equation}
\bigg[\nabla^2 + k^2 - \frac{2\mu}{\hbar^2} U_{c}({\bf R}) + U_{so}({\bf R}) {\mathbf{\sigma \cdot L}}\bigg] \chi({\bf k},{\bf r}) = 0,
\label{sch1}
\end{equation}
where $U_c$ and $U_{so}$ are the central and spin-orbit contributions to the neutron-nucleus and proton-nucleus microscopic optical 
potentials derived in our model \cite{lopez21}, and in the case of protons also includes the Coulomb potential. These nucleon-nucleus 
potentials can be written in terms of the isoscalar and isovector potentials, $U_0$ and $U_1$ respectively, by inverting Eqs.(\ref{ompot0}) and 
(\ref{ompot1}).

It is useful to perform the partial wave decomposition of the solutions of (\ref{sch1}) as
\begin{equation}
\chi = 4\pi \sum_{JlM} i^l \frac{u_{Jl}(kr)}{kr} {\cal Y}_{ls}^{{JM}^*}(\hat{\bf k}) {\cal Y}_{ls}^{JM}(\hat{\bf r}),
\label{partiald}
\end{equation}
where the vector spherical harmonics defined as
\begin{equation}
{\cal Y}_{ls}^{JM}= \sum_{m,\nu} \langle l m s \nu \vert JM \rangle Y_{lm} \vert s \nu \rangle,
\label{vsh}
\end{equation}
with $J= l \pm 1/2$ the total angular momentum resulting from the coupling between the spin and orbital angular momenta. The radial
functions $u_{Jl}$ are the solutions of the radial Sch\"odinger equation
\begin{equation}
\frac{d^2 u_{Jl}}{d \rho^2}+\left\{1 - \left[\frac{U_{c}(\rho)}{E} + 
%\left(\begin{array}{c}l\\ o \\-l-1\end{array}\right)
[J(J+1)-l(l+1)-\frac{3}{4}] \frac{U_{so}(\rho)}{E}\right]-\frac{l(l+1)}{\rho^2}\right\}u_{Jl}=0.
\label{sch2}
\end{equation}
where $\rho=kr$.

The spin-averaged moduli of the distorted wave functions for incoming protons and outgoing neutrons can be obtained by using its partial waves 
decomposition (\ref{partiald}) as \cite{danielewicz17}
\begin{eqnarray}
\overline{|\chi(\vec{r})|^2}&=&\frac{1}{2s+1}\sum_L(2L+1)A_L^{\rho}(r)P_L(\cos\theta) \\
A_L^{\rho}&=&\sum_{J'l'}(2J'+1)(2l'+1)\sum_{Jl}(2J+1)(2l+1) \nonumber \\
&\times&\left(\begin{array}{ccc} l & l' & L \\ 0  & 0  & 0\end{array}\right)^2\left\{\begin{array}{ccc} l & l' & L \\ J' & J & s\end{array}
\right\}^2 \frac{1}{k^2r^2} Re\left[i^{l-l'} u^*_{J'l'}(r) u_{Jl}(r) \right],
\label{modulus}
\end{eqnarray}
where the functions $u_{Jl}$ are the solutions of (\ref{sch2}) for each partial wave.

Using again the partial wave decomposition, the DWBA cross section for the $(p,n)$ charge exchange reactions can be written as \cite{danielewicz17}  
\begin{eqnarray}
\frac{d\sigma_{(p,n)}}{d\Omega}&=&\frac{1}{k_p^2}\frac{1}{2s+1}\sum_L(2L+1)A_L^{(p,n)}P_L(\cos\theta) \\
A_L^{(p,n)}&=&4\mu_p\mu_nk_pk_n\sum_{J'l'}(2J'+1)(2l'+1)\sum_{Jl}(2J+1)(2l+1) \nonumber \\
&\times&\left(\begin{array}{ccc} l & l' & L \\ 0  & 0  & 0\end{array}\right)^2\left\{\begin{array}{ccc} l & l' & L \\ J' & J & s\end{array} 
\right\}^2 Re\left[I^*_{J'l'}I_{Jl}\right].
%\\ \nonumber 
\label{cspn}  
\end{eqnarray}

The functions $I_{Jl}$ are the radial matrix elements of (\ref{transition}), which read 
\begin{equation}
I_{Jl}=2\frac{\sqrt{N-Z}}{A}\int_0^\infty dr\; u_{nJl}^{(+)}(r)U_1^{Jl}(r)u_{pJl}^{(+)}(r) 
\label{rmatel}
\end{equation}

The spin-orbit contributes to the isovector part of the optical potential and has an impact on the analyzing power, which can be understood as follows 
\cite{gosset76}. Within the Lane model it is assumed that quasielastic exchange reactions $(p,n)$ do not modify the total, orbital and spin angular 
momenta between entrance and exit channels. If in addition it is also assumed a zero spin target nucleus, there are only two possible amplitudes 
that correspond to the no-flip and flip of the spin projection of the projectile. The analyzing power is related to these amplitudes by
\begin{equation}
A = \frac{2 Img (Y X^*)}{\vert X \vert^2 + \vert Y \vert^2},
\label{anpow1}
\end{equation}
where the amplitudes $X$ and $Y$ are given by \cite{gosset76,satchler64a} 
\begin{equation}
X= \frac{4 \pi}{k_n k_p}\sum_{l=0}^{\infty} \big[(l+1) I_{J=l+1/2, l} + l I_{J=l-1/2, l}\big]P_l(cos \theta)
\label{anpow2x}
\end{equation}
\begin{equation}
Y= \frac{4 \pi}{k_n k_p}\sum_{l=0}^{\infty} \big[I_{J=l+1/2, l} - I_{J=l-1/2, l}\big]P^1_l(cos \theta),
\label{anpow2y}
\end{equation}
where $P_l(cos \theta)$ and $P^1_l(cos \theta)$ are the Legendre polynomials and associated Legendre polynomials with $m=1$, respectively.

\section{Results and discussions}

\begin{figure}[ht]
\hspace*{2cm}
\centering
\includegraphics[width=0.75\linewidth,clip=true]{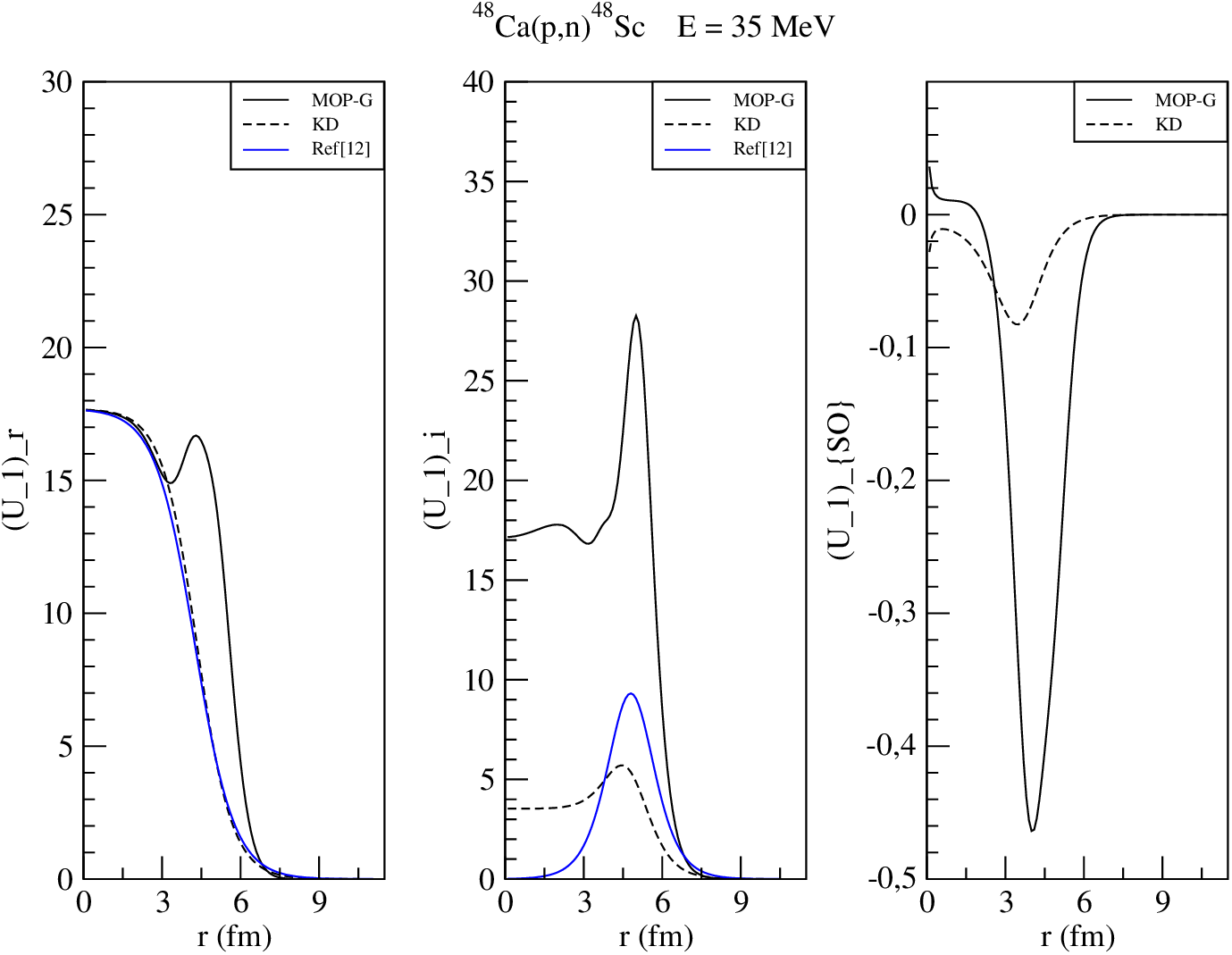}
\caption{Real and imaginary parts of the central potential and spin-orbit term of the isovector potential
$U_1$ in the charge exchange reaction $^{48}$Ca(p,n)$^{48}$Sc at 35 MeV computed with the MOP-G 
and KD models. The real and imaginary parts of the central term of the isovector potential fitted 
in Ref.\cite{patterson75} to charge exchange reaction data \cite{doering75} is also displayed.}  
\label{fig:fig39}
\end{figure}

\begin{figure}[ht]
\hspace*{2cm}
\centering
\includegraphics[width=0.75\linewidth,clip=true]{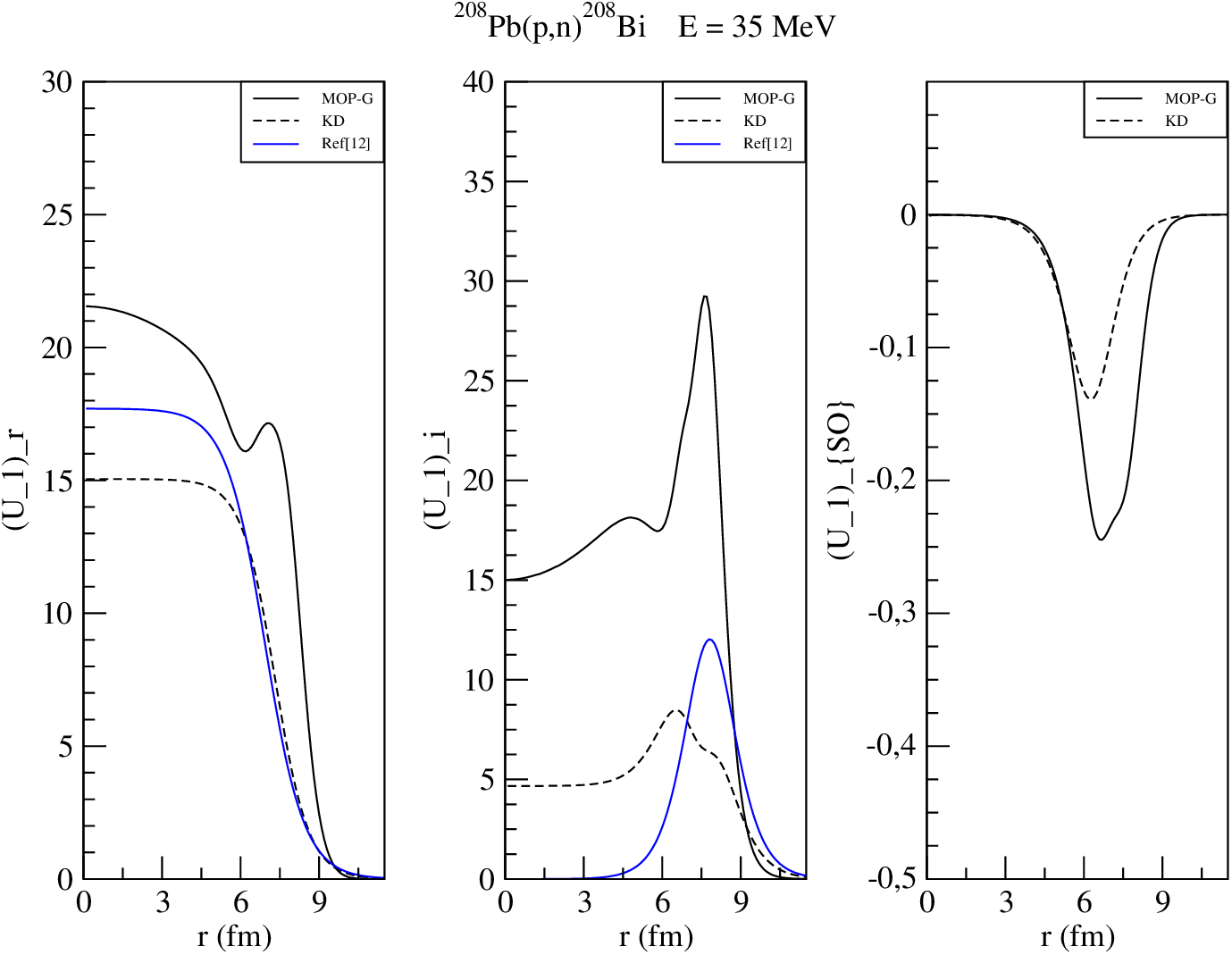}
\caption{The same as in Fig.\ref{fig:fig39} but for the reaction $^{208}$Pb(p,n)$^{208}$Bi at 45 MeV}
\label{fig:fig39a}
\end{figure}

The isovector potential used in this work is provided by Eq.(\ref{ompot1}) computed with $U_p({\bf R})$ and $U_n({\bf R})$ obtained with our
theoretical MOP-G and with the phenomenological optical potential of Koning and Delaroche (KD) \cite{koning03} that will be used for useful
comparisons. For the following discussions is useful to keep in mind that the MOP-G is not fitted to any nuclear reaction data 
\cite{lopez21} and that the parameters of the KD model are fitted to nucleon-nucleus elastic scattering data. Thus these two models allow to 
check how much optical models fitted in another scenarios are also able to describe charge exchange reactions.
In Figure \ref{fig:fig39} we display the real and imaginary parts of the central isovector potential as well as the spin-orbit 
contribution to the isovector potential for the $^{48}$Ca(p,n)$^{48}$Sc reaction obtained using the MOP-G and KD models. In the same
figure we also display the real and imaginary parts of the central term of the phenomenological optical potential fitted in Ref.~\cite{patterson75} 
to reproduce the charge exchange $(p,n)$ measured data for incident protons of 25, 35 and 45 MeV on $^{48}$Ca, $^{90}$Zr, $^{120}$Sn and 
$^{208}$Pb targets \cite{doering75}. 
To compute the MOP-G the necessary input are the neutron and proton densities \cite{lopez21}. In this reference these densities are obtained by a 
HF calculation using the quasi-local energy density functional theory, which allows to deal with HF calculations in finite nuclei with finite-range
forces in coordinate space (see for more details Appendix B of \cite{lopez21}). However in this work we use parametrized densities of Fermi type 
that minimize the semiclassical Extended Thomas-Fermi energy density functional built up with the Gogny D1S force (see \cite{bhagwat21} for further 
details). The real part of the isovector potential predicted by the MOP-G model has a central strength of about 17 MeV, which decreases at the 
surface. The imaginary part has a similar strength in the center but is strongly peaked at the surface, pointing out that within the MOP-G model 
the surface absorption is important. The central isovector potential computed with the KD model predicts a similar real part but a smaller 
imaginary contribution. This is a direct consequence of the fact that the imaginary part of the nucleon-nucleus optical potential computed with KD 
is smaller that the one obtained with the MOP-G model, as it can be seen in Figures 1 and 2 of \cite{lopez21}. The spin-orbit form factor in the 
isovector potential is obtained in the MOP-G model as the difference between the neutron and proton spin-orbit form factors computed 
in the residual nucleus and in the target, respectively as explained in \cite{lopez21}, while in the KD model the isovector spin-orbit
term is obtained from the spin-orbit potentials in the exit and entrance channels. The central real part predicted by the MOP-G
and KD models is similar to the one predicted by the phenomenological optical model of \cite{patterson75}, while the central imaginary contribution 
of \cite{patterson75} is also peaked and lies midway between the predictions of MOP-G and KD models. We have also computed the isovector potentials
$U_1$ for the same $^{48}$Ca(p,n)$^{48}$Sc reaction but at energies of 25 and 45 MeV. These calculations reveal that the real part of $U_1$
computed with MOP-G is almost independent of the energy of the projectile, in agreement with the real part of $U_1$ fitted in \cite{patterson75}, 
whereas the prediction of KD has a stronger energy dependence, decreasing its strength with growing energy of the projectile. Also the imaginary 
part of $U_1$ increases the volume absorption and decreases the surface one when the energy of the projectile grows, which is in agreement with the 
decreasing trend of the phenomenological isovector potential of \cite{patterson75}. To get more insight about the dependence on the size of the 
target, we display in Figure \ref{fig:fig39a} the real and imaginary parts of the central term of the isovector potential $U_1$ as well as the 
spin-orbit contribution to this potential using the same models as in Figure \ref{fig:fig39} but in the case of the $^{208}$Pb(p,n)$^{208}$Bi reaction
at 35 MeV. As a function of the energy of the projectile the behaviour of the different isovector potentials displayed in Figure \ref{fig:fig39} 
is the same than the previously discussed for the $^{48}$Ca(p,n)$^{48}Sc$ reaction. The size of the real and imaginary parts of the isoscalar 
potential scales following a $A^{1/3}$ law. The real contribution of MOP-G to $U_1$ is deeper for the $^{208}$Pb target than for the $^{48}$Ca one, 
while the contrary happens for the KD prediction. The imaginary part grows when the mass of the target increases, in particular the volume 
contribution for both MOP-G and KD models, which is in agreement with the behaviour of the imaginary part in the case of the phenomenological 
potential \cite{patterson75}. 
 
\begin{figure}[ht]
\hspace*{2cm}
\centering
\includegraphics[width=0.6\linewidth,clip=true]{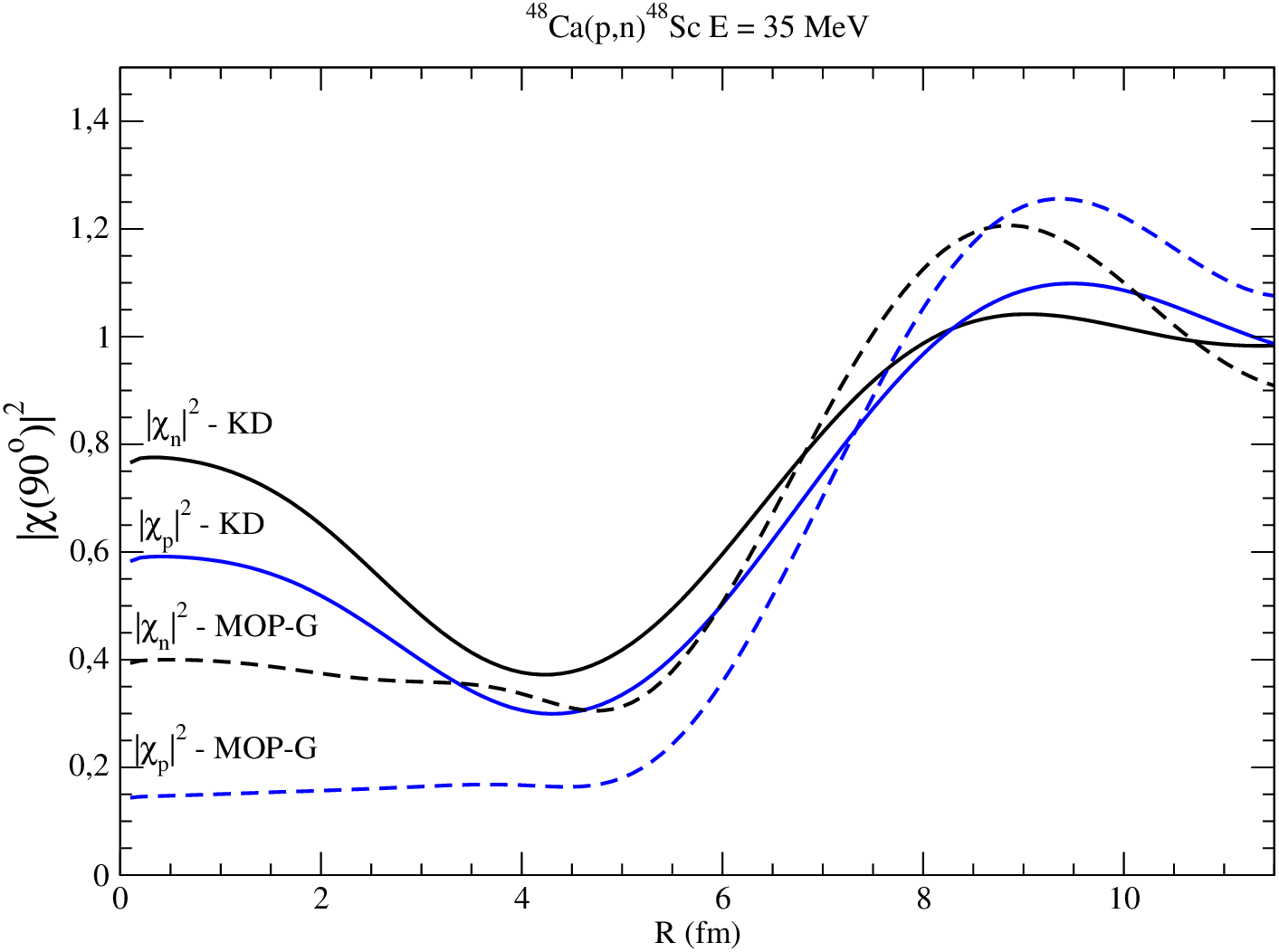}
\caption{Spin-averaged square moduli of the distorted waves corresponding to the initial and final states in the direction 
perpendicular to the initial and final wave vectors of the of the quasielsastic charge exchange reaction $^{48}$Ca(p,n)$^{48}$Sc 
as a function of the distance from the center of the target computed with the MOP-G and KD models.}
\label{fig:fig38}
\end{figure}

The square of the modulus of the wave functions in the entrance and exit channels computed with the MOP-G and KD models is displayed in Figure 
\ref{fig:fig38} in the case of the $^{48}$Ca(p,n)$^{48}$Sc reaction.  We see that at large distances the neutron and proton wave functions computed
with both models oscillate around one as expected. At distances of the order of the nuclear size both moduli show a depletion, which is due to the
loss of probability flux to other channels that is described by the imaginary part of the optical potential \cite{danielewicz17}. We can also see that
in the case of KD-model the moduli increase again around the center of the nucleus due to constructive interferences \cite{danielewicz17}, while 
this effect is absent in the case of the MOP-G model probably due to the stronger imaginary part of the isovector potential.

\begin{figure}[ht]
\hspace*{-1cm}
\centering
\includegraphics[width=0.8\linewidth,clip=true]{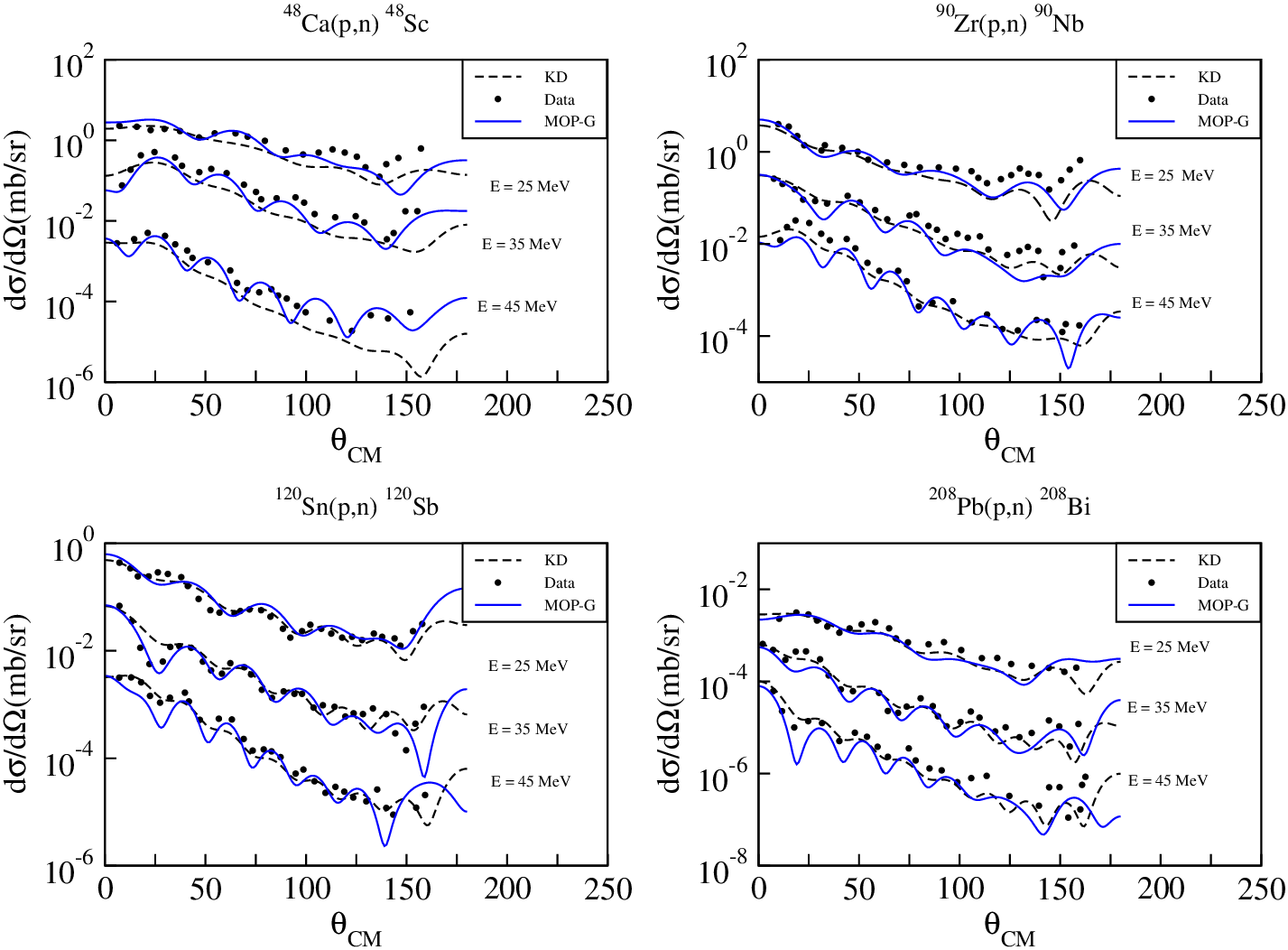}
\caption{Differential cross sections of the $^{48}$Ca(p,n)$^{48}$Sc,$^{90}$Zr(p,n)$^{90}$Nb, $^{120}$Sn(p,n)$^{120}$Sb and 
$^{208}$Pb(p,n)$^{208}$Bi reactions at incident proton energies 45, 35 and 25 MeV predicted by the MOP-G and KD models 
compared to the experimental data of Ref.\cite{patterson75}}.
\label{fig:fig40}
\end{figure}

\begin{figure}[ht]
\hspace*{-1cm}
\centering
\includegraphics[width=0.8\linewidth,clip=true]{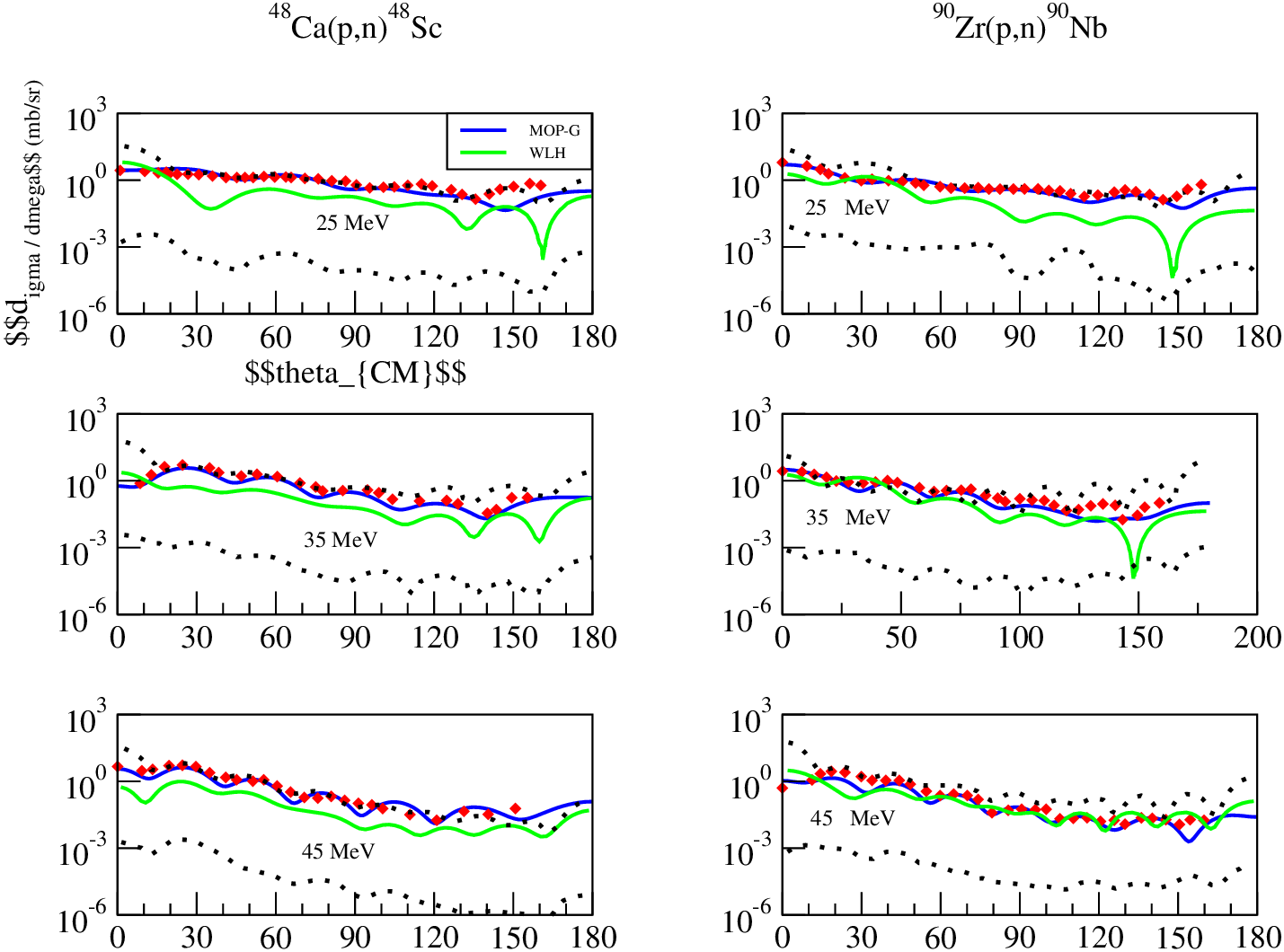}
\caption{Differential cross sections of the $^{48}$Ca(p,n)$^{48}$Sc and $^{90}$Zr(p,n)$^{90}$Nb reactions at incident proton energies 
of 25, 35 and 45 MeV predicted by the MOP-G (blue lines) and WLH (green lines) models compared to the experimental data of Ref.\cite{patterson75}
(red symbols). Dotted lines indicate the 95\% confidence interval of the WLH microscopic optical potential.}
\label{fig_wlh}
\end{figure}

\begin{figure}[ht]
\hspace*{-1cm}
\centering
\includegraphics[width=1.2\linewidth,clip=true]{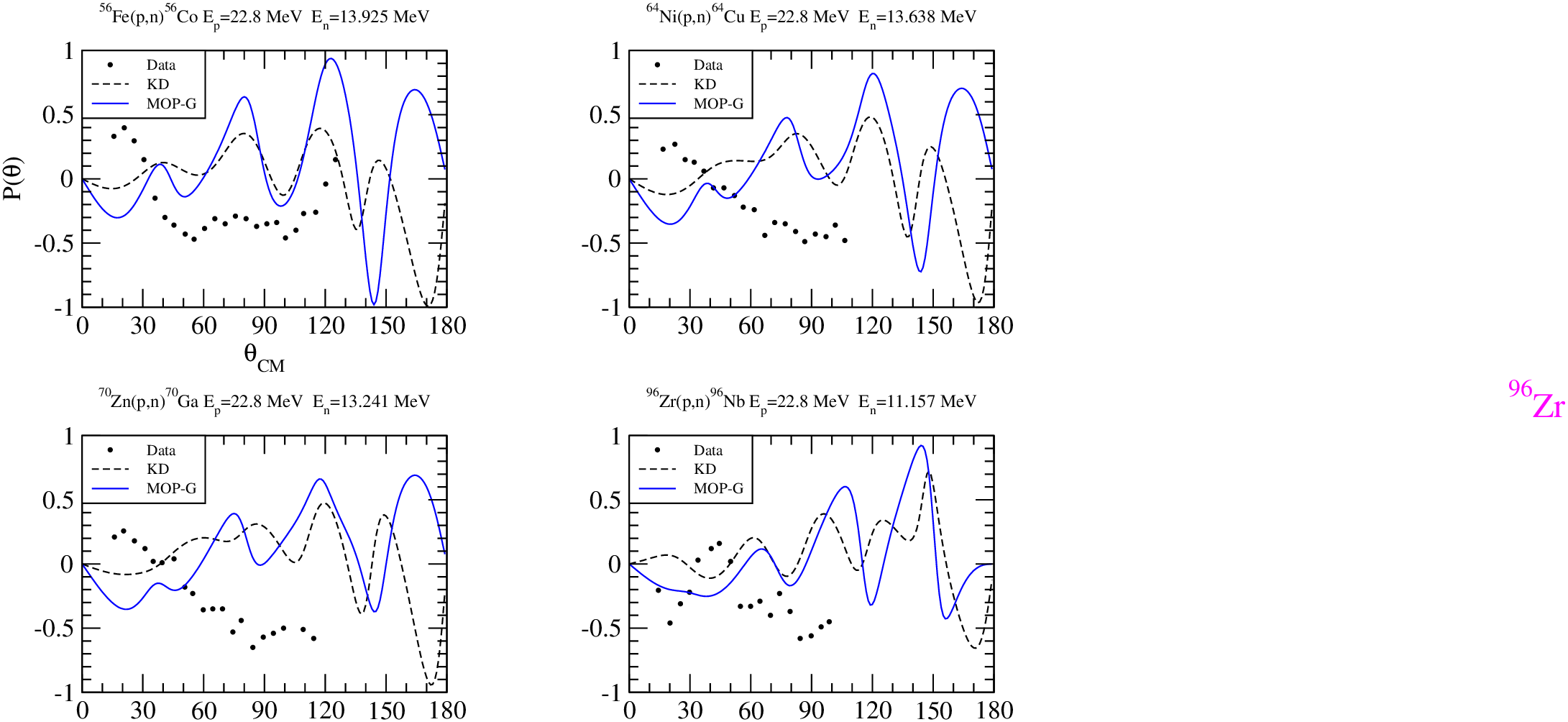}
\caption{Theoretical analyzing power of quasielastic charge exchange reactions $^{56}$Fe(p,n)$^{56}$Co, $^{64}$Ni(p,n)$^{64}$Cu,
$^{70}$Zn(p,n)$^{70}$Ga and $^{96}$Zr(p,n)$^{96}$Nb at incident proton energies of 22.8 MeV computed with the MOP-G and KD models
and compare with the experimental values of Ref.\cite{gosset76}}
\label{fig:fig41}
\end{figure}

In order to study the isovector part of our MOP-G we display in the four panels of Figure \ref{fig:fig40} the differential cross sections for the 
quasi-elastic charge exchange reactions $^{48}$Ca(p,n)$^{48}$Sc,$^{90}$Zr(p,n)$^{90}$Nb, $^{120}$Sn(p,n)$^{120}$Sb and $^{208}$Pb(p,n)$^{208}$Bi 
from the ground-state to the corresponding IAS at proton energies of 25, 35 and 45 MeV computed with the MOP-G and compare with the experimental 
values \cite{patterson75}. The angular distributions for the same charge exchange reactions predicted by the KD optical 
potential, obtained using again the formalism of Section 2, are also shown in the same figure for useful comparisons. The differential cross 
sections for both models have been calculated using Eq.(\ref{cspn}). However, we have also 
checked that the same result can be obtained from Eq.(\ref{cs}) with $|T_{pn}|^2=|X|^2+|Y|^2$ with $X$ and $Y$ given by  
Eqs.(\ref{anpow2x}) and (\ref{anpow2y}), respectively. We can see that the differential cross sections predicted by the MOP-G model (solid lines 
in the figure) are in quite good agreement with the experimental values for all the targets and all the energies considered, following rather well 
the decreasing and oscillatory trend with the scattering angle shown by the experimental data. These results point out that the isovector part of 
the MOP-G built up within the Lane model is very reasonable. In the analyzed reactions the KD results (dashed lines in the figure) decrease rather
 smoothly with the scattering angle with more damped oscillations and, in general, averaging the MOP-G cross sections. In spite of the fact that the 
MOP-G model is free of parameters fitted to scattering data, we see that the predictions for the considered reactions also compare rather well with 
other theoretical results reported in earlier literature such us the ones displayed in Figure 7 of \cite{bauge01} and Figures 2 and 3 of 
\cite{koha07} as well with the phenomenological fits of the experimental data of Ref.\cite{patterson75} and \cite{danielewicz17}.

In Figure \ref{fig_wlh} we compare the differential cross sections for the quasielastic $^{48}$Ca(p,n)$^{48}$Sc and $^{90}$Zr(p,n)$^{90}$Nb reactions
 at incident energies of 25, 35 and 45 MeV predicted by our MOP-G model with the same results reported in Ref.\cite{whitehead22}, which are computed 
with the MOP derived by Whitehead, Lim and Holt (WLM) from many-body perturbation theory with chiral forces \cite{whitehead21}. We see that the 
differential cross sections computed with the MOP-G model are slightly larger and lie closer to the experimental values that the predictions of the 
WLH calculation. In the same figure we also show the 95\% confidence interval of the WLM MOP due to the underlying uncertainties of the chiral forces, 
which is very similar to the same confidence interval obtained by a Bayesian analysis in the case of the phenomenological KD approach 
\cite{whitehead22}.

In order to analyze the influence of the spin-orbit part of the optical potential on the quasi-elastic $(p,n)$ reactions we display in 
Figure \ref{fig:fig41} the analyzing power of the reactions $^{56}$Fe(p,n)$^{56}$Co, $^{64}$Ni(p,n)$^{64}$Cu, $^{70}$Zn(p,n)$^{70}$Ga and 
$^{96}$Zr(p,n)$^{96}$Nb  for incident protons of 22.8 MeV computed with the MOP-G and KD models. The experimental data are taken from 
Ref.\cite{gosset76}. We see that the predicted analyzing power by these two theoretical models show an oscillatory trend, which is 
more marked for the MOP-G model. The experimental analyzing powers for the different reactions considered here also show, roughly, 
an oscillatory structure, which is more clear in the $^{56}$Fe(p,n)$^{56}$Co and $^{96}$Zr(p,n)$^{96}$Nb reactions. We see that globally the 
 analyzing powers predicted by the MOP-G and KD models are, on the one hand, shifted towards positive values, and, on the other hand, the 
oscillations are enhanced and out of phase as compared to the experimental data. Overall the detailed behaviour of the experimental analyzing power
 for the considered reactions described by the MOP-G and KD models is  not reproduced. Although the analyzing powers of 
the same reactions displayed in Figure 6 of Ref.\cite{gosset76} and in panel (c) of Ref.\cite{bauge01} reproduce better the experimental data 
than the predictions of the MOP-G and KD models, it shall be pointed out that, on the one hand, the optical models used in \cite{gosset76} and 
\cite{bauge01} are fitted to experimental data of charge exchange reactions, and, on the other hand, the agreement between the predictions of the
models reported in  Refs.\cite{gosset76,bauge01} with the experimental data of the analyzing power is not extremely precise. 
\begin{table}[ht] 
\begin{center}  
\caption{Total cross sections in \textit{mbarns} for the charge exchange reactions studied in figure \ref{fig:fig40}. Theoretical calculations 
performed with the MOP-G and KD models are compared with the experimental values taken from \cite{doering75} and with the predictions of the 
phenomenological model of Ref.~\cite{patterson75}} 
\scalebox{1.0}{ \begin{tabular}{|ccccc|ccccc|} \multicolumn{5}{c}{$^{48}Ca(p,n) ^{48}Sc$} & 
\multicolumn{5}{c}{$^{90}Zr(p,n) ^{90}Nb$}\\ \hline E & KD & MOP-G & \cite{patterson75} & EXP & E & KD & MOP-G & \cite{patterson75} & EXP\\ 45 
& 4.38 & 6.09 & 6.90 & 8.4 $\pm$ 1.0 & 45 & 2.79 & 2.28 & 3.57 & 4.4 $\pm$ 0.5 \\ 35 & 5.96 & 7.45 & 9.15 & 10.2 $\pm$ 1.1 & 35 & 3.96 & 3.36 
& 4.60 & 4.8 $\pm$ 0.5 \\ 25 & 7.92 & 10.92 & 11.22 & 10.6 $\pm$ 1.2 & 25 & 5.54 & 6.21 & 5.70 & 6.7 $\pm$ 0.8 \\ \hline 
\multicolumn{5}{c}{$^{120}Sn(p,n) ^{120}Sb$} & \multicolumn{5}{c}{$^{208}Pb(p,n) ^{208}Bi$}\\ \hline 45 & 4.14 & 2.98 & 4.60 & 5.8 $\pm$ 0.7 & 
45 & 4.34 & 2.57 & 4.40 & 5.4 $\pm$ 0.6 \\ 35 & 5.94 & 4.62 & 6.00 & 5.6 $\pm$ 0.6 & 35 & 6.52 & 4.86 & 5.80 & 6.8 $\pm$ 0.8 \\ 25 & 8.28 & 
8.97 & 7.54 & 8.5 $\pm$ 1.0 & 25 & 8.66 & 8.62 & 6.40 & 9.6 $\pm$ 1.1 
\\ \hline 
\end{tabular}} 
\label{table:table12}

\end{center}
\end{table}

The reaction cross sections for the  $^{48}$Ca(p,n)$^{48}$Sc,$^{90}$Zr(p,n)$^{90}$Nb, $^{120}$Sn(p,n)$^{120}$Sb and $^{208}$Pb(p,n)$^{208}$Bi
reactions at incident proton energies of 45, 35 and 25 MeV computed by the MOP-G and KD models are given in Table \ref{table:table12} 
together with the measured values reported in \cite{doering75} and the results predicted by the optical model of
Ref.\cite{patterson75}, which was fitted to reproduce these experimental data. For a given target the 
theoretical cross sections decrease with increasing bombarding energy for all considered reactions. This trend is in agreement with 
the behaviour shown by the experimental data and by the total cross sections computed with the optical potential \cite{patterson75}. 
 We see that for the lowest considered energy, i.e. 25 MeV, the MOP-G and KD predictions lie within the experimental window data, while for 
higher energy the theoretical predictions underestimate, in general, the experimental values. We see that as a function of the energy
the cross sections predicted  by the MOP-G and KD models decrease faster as compared to the values predicted by the optical model of 
Ref.\cite{patterson75}, which seems to point out some deficiencies in the energy dependence of the MOP-G and KD models for describing accurately 
the cross sections of these exchange reactions at high energies.  

\section{Conclusions}
We have examined the isovector properties of our semi-microscopic nucleon-nucleus optical potential derived in a previous work, which is based on a 
nuclear matter approach with a Brueckner-Hartree-Fock calculation using a G-matrix built up with an effective Gogny interaction fitted
to describe ground-state properties of finite nuclei. Within a Lane prescription we split the optical potential in its isoscalar and isovector 
parts. The transition matrix elements are evaluated using the distorted
Born wave functions in the entrance and exit channels. These matrix elements allow to compute the differential cross sections and the analyzing
powers as a function of the scattering angle. We find that our semi-microscopic model, which does not contain adjustable parameters fitted to
scattering data, reproduce the experimental data of quasi-elastic $(p,n)$ reactions for the considered reactions in a remarkably good agreement
with the experimental data. Also the description provided by our model is fully comparable with results obtained using other different theoretical
optical potentials, in particular with the recently reported results provided by the microscopic optical potential derived by Whitehead, Lim and Holt 
using chiral forces. However, the analyzing powers predicted by our model and by the KD global optical potential are poor as compared 
with the experimental data and the theoretical predictions of other optical models specially fitted to reproduce the experimental behaviour of this
observable. The reaction cross sections predicted by 
the MOP-G and KD  models reproduce the experimental data for the lowest bombarding energy considered in this 
work, i.e. 25 MeV, but both models fail at higher energies, 35 and 45 MeV, underestimating the measured reaction cross sections. We conclude that 
the isovector part of the semi-microscopic optical potential based on the Gogny force is suitable to describe in a reasonable way
scattering scenarios where the isovector part of the optical potential plays a relevant role. The poor description of the 
analyzing powers and the too strong decreasing of the total cross sections in quasielastic charge exchange reactions point out the limits 
of optical models, such as  MOP-G and KD, fitted in another scenarios different from charge exchange reactions. A better description of the 
experimental data of charge exchange reactions by the MOP-G and KD models may be obtained, for example, introducing suitable renormalization factors
 of the real and imaginary parts of the isovector potential, as it is done in double-folding optical potentials \cite{koha07}. However, a word of 
caution should also be given. The Lane formalism using optical potentials fitted to experimental data of charge exchange reactions together with 
the DWBA can reproduce well the differential cross sections and in a reasonably way the analyzing power and the energy dependence of the cross 
sections as a function of the energy as it can be seen in Refs.~\cite{gosset76,bauge01}. However, a more accurate description of these charge 
exchange reactions would require more sophisticated method beyond the simple Lane model.
\section*{Acknowledgement}
One of the authors (X.V.) acknowledges the partial support from 
Grants No. PID2020-118758GB-I00
and No. CEX2019-000918-M (through the “Unit of
Excellence María de Maeztu 2020-2023” award to ICCUB)
from the Spanish MCIN/AEI (DOI 10.13039/501100011033).
The authors are indebted to J.N.De for useful discussions.

\section*{Data availability}

The data that supports the findings of this study are available
upon reasonable request from the authors.

\end{document}